\documentclass[prd,twocolumn,superscriptaddress,nofootinbib,showpacs]{revtex4}

\usepackage{graphicx,amsmath,amsfonts,amssymb,revsymb,dcolumn,epsfig,bm}

\begin{document}
\title{Lookback time bounds from energy conditions}

\author{J. Santos} \email{janilo@dfte.ufrn.br}
\affiliation{Universidade Federal do Rio Grande do Norte, 
Departamento de F\'{\i}sica C.P. 1641, 59072-970 Natal -- RN, Brasil}
\affiliation{Departamento de Astronomia, Observat\'orio Nacional, 
20921-400, Rio de Janeiro -- RJ, Brasil}
\affiliation{Centro Brasileiro de Pesquisas F\'{\i}sicas, 
22290-180 Rio de Janeiro -- RJ, Brasil}

\author{J.S. Alcaniz}\email{alcaniz@on.br}
\affiliation{Departamento de Astronomia, Observat\'orio Nacional, 
20921-400, Rio de Janeiro -- RJ, Brasil}

\author{M.J. Rebou\c{c}as}\email{reboucas@cbpf.br}
\affiliation{Centro Brasileiro de Pesquisas F\'{\i}sicas, 
22290-180 Rio de Janeiro -- RJ, Brasil}

\author{N. Pires}\email{npires@dfte.ufrn.br}
\affiliation{Universidade Federal do Rio Grande do Norte,  
Departamento de F\'{\i}sica C.P. 1641, 59072-970 Natal -- RN, Brasil}

\date{\today}

\begin{abstract}
In general relativity, the energy conditions are invoked to restrict general 
energy-momentum tensors on physical grounds. We show that in 
the  standard Friedmann--Lema\^{\i}tre--Robertson--Walker (FLRW) approach to 
cosmological modeling, where the energy and matter components of the cosmic 
fluid are unknown, the energy conditions provide model-independent 
bounds on the behavior of the lookback time of cosmic sources as a function of 
the redshift for any value of the spatial curvature. We derive and confront such bounds 
with a lookback time sample which is built from the age estimates of 32 galaxies 
lying in the interval $0.11 \lesssim z \lesssim 1.84$ and by assuming the 
total expanding age of the Universe to be $13.7 \pm 0.2$ Gyr, as obtained 
from current cosmic microwave background experiments. In agreement with 
previous results, we show that all energy conditions seem to have been 
violated at some point of the recent past of cosmic evolution.
\end{abstract}


\maketitle

\section{Introduction}

The standard Friedmann-Lema\^{\i}tre-Robertson-Walker (FLRW) 
approach to model the Universe begins with two basic assumptions.  
First, it is assumed that our $3$--dimensional space is homogeneous 
and isotropic at sufficiently large scales. Second, it is also assumed 
the Weyl's principle, that ensures the existence of a cosmic time $t$. 
The most general spacetime metric consistent with these assumptions is 
\begin{equation}
\label{RWmetric}
ds^2 =  dt^2 - a^2 (t) \left[\, \frac{dr^2}{1-kr^2} + r^2(d\theta^2 
         + \sin^2 \theta  d\phi^2) \,\right],
\end{equation}
where ($k=0,1,-1$), $a(t)$ is the cosmological scale factor, and we 
have set the speed of light $c = 1$. The metric~(\ref{RWmetric}) only 
expresses the above assumptions, and to proceed further in this geometrical 
approach to model the physical world, one needs a metrical theory of 
gravitation as, e.g., general relativity  (which we assume in this work) 
to study the dynamics of the Universe.

These very general assumptions imply that the cosmological fluid  
is necessarily of a perfect-fluid type, i.e.,
\begin{equation} \label{EM-Tensor}
T_{\mu\nu} = (\rho+p)\,v_\mu v_\nu - p \,g_{\mu \nu}\;,
\end{equation}
where $v_\mu$ is the fluid four-velocity, with total density $\rho$ 
and pressure $p$ given, respectively, by 
\begin{eqnarray}
\rho & = & \frac{3}{8\pi G}\left[\,\frac{\dot{a}^2}{a^2}
                                  +\frac{k}{a^2} \,\right]\;,
\label{rho-eq} \\
p & = & - \frac{1}{8\pi G}\left[\, 2\,\frac{\ddot{a}}{a} +
\frac{\dot{a}^2}{a^2} + \frac{k}{a^2} \,\right] \;, \label{p-eq}
\end{eqnarray}
and  dots denote derivative with respect to the time $t$.

Without assuming any particular equation of state, one can proceed 
even further in this approach to model the universe by invoking the 
so-called \emph{energy conditions\/}~\cite{Hawking-Ellis,Visser,Carroll}  
that limit the energy-momentum tensor on physical grounds.

In the FLRW context, where only the energy-momentum of a perfect 
fluid~(\ref{EM-Tensor}) should be considered, these conditions take one 
of the forms (see, e.g., \cite{Hawking-Ellis,Visser,Carroll,Visser-Barcelo})
\begin{equation} \label{ec}
\begin{array}{llll}
\mbox{\bf NEC} \ & \Longrightarrow &\, \rho + p \geq 0 \;,  &   \\
\\
\mbox{\bf WEC} \ & \Longrightarrow & \rho \geq 0 &
\ \mbox{and} \quad\, \rho + p \geq 0 \;,  \\
\\
\mbox{\bf SEC}   &\Longrightarrow & \rho + 3p \geq 0 &
\ \mbox{and} \quad\, \rho + p \geq 0 \;, \\
\\
\mbox{\bf DEC}   &\Longrightarrow & \rho \geq 0  &
\ \mbox{and} \; -\rho \leq p \leq\rho \;,
\end{array}
\end{equation}
where NEC, WEC, SEC and DEC correspond, respectively, to the null, weak, 
strong and dominant energy conditions. From  Eqs.~(\ref{rho-eq})~--~(\ref{p-eq}), 
one has that these energy conditions can be recast as a set of differential 
constraints involving the scale factor $a$ and its derivatives for any 
spatial curvature $k\,$, namely 
\begin{eqnarray}
\mbox{\bf NEC} & \, \Longrightarrow & \quad\, - \frac{\ddot{a}}{a}
+  \frac{\dot{a}^2}{a^2}  + \frac{k}{a^2} \geq 0 \;, \label{nec-eq}
\\
\mbox{\bf WEC} & \, \Longrightarrow & \quad\;\; \frac{\dot{a}^2}{a^ 2} 
+ \frac{k}{a^2} \geq 0 \;,
\label{wec-eq}
\\
\mbox{\bf SEC} & \, \Longrightarrow & \quad\;\; \frac{\ddot{a}}{a} \leq 0 \;,
\label{sec-eq} \\
\mbox{\bf DEC} & \, \Longrightarrow & \; - 2\left[
\frac{\dot{a}^2}{a^2}+\frac{k}{a^2} \right] \leq
\frac{\ddot{a}}{a}  \leq
\frac{\dot{a}^2}{a^2}+\frac{k}{a^2} \;. \label{dec-eq}
\end{eqnarray}
where from the above NEC/WEC inequations~(\ref{ec}), the NEC restriction 
[Eq.~(\ref{nec-eq})] is also part of both  WEC and NEC constraints.

A further step in this FLRW approach is to confront the energy condition 
predictions with the observations. In this regard, since the pioneering works 
by Visser~\cite{M_Visser1997}, it has been shown that in the FLRW framework  
the energy conditions provide model-independent bounds on the cosmological 
observables, and an increasing number of analyses involving such bounds 
have been discussed in the recent literature. 
In Ref.~\cite{SAR2006,SAPR2007}, e.g., we have studied the behavior of the 
distance modulus to cosmic sources for any spatial curvature and 
confronted these bound predictions with current type Ia  supernovae 
(SNe Ia) observations. 
In Ref.~\cite{Sen_Scherrer2007}, by assuming a flat ($k=0)$ universe  and 
that only the dark energy component obeys the WEC, it was derived upper 
limits on the matter density parameter $\Omega_m$. More recently,  
the confrontation of the NEC and SEC bounds with a combined sample 
of 192 supernovae was carried  out providing similar 
and complementary results~\cite{Gong-et-al2007,Gong_Wang2007}. 
Energy conditions constraints on alternative gravity models, such as the 
so-called $f(R)$-gravity, have also been investigated in Ref.~\cite{santiago} 
(see also\cite{EC1} for other analyses).

In this paper, to shed some additional light on the interrelation between 
energy conditions and observational data, we extend and complement our previous
results~\cite{SAPR2007,SAR2006} in two different ways. First, we derive 
model-independent energy-conditions bounds on the lookback time-redshift 
relation $t_L(z)$  in an expanding FLRW  universe for any spatial curvature $k$. 
Second, by assuming the total age of the Universe to be $t_0 = 13.7 \pm 0.2$ Gyr, 
as given by current Cosmic Microwave Background (CMB) measurement~\cite{WMAP}, 
we transform the age estimates of 32 galaxies, as provided in Ref.~\cite{Simon_Verde_Jimenez2005}, into lookback time observations and 
concretely confront these data with the energy conditions bound predictions. 
Similarly to our previous studies, we find that all energy conditions seem to 
have been violated at some point of the cosmic evolution.

\begin{figure*}[t]
\includegraphics[width=6.0cm,height=8.0cm,angle=270]{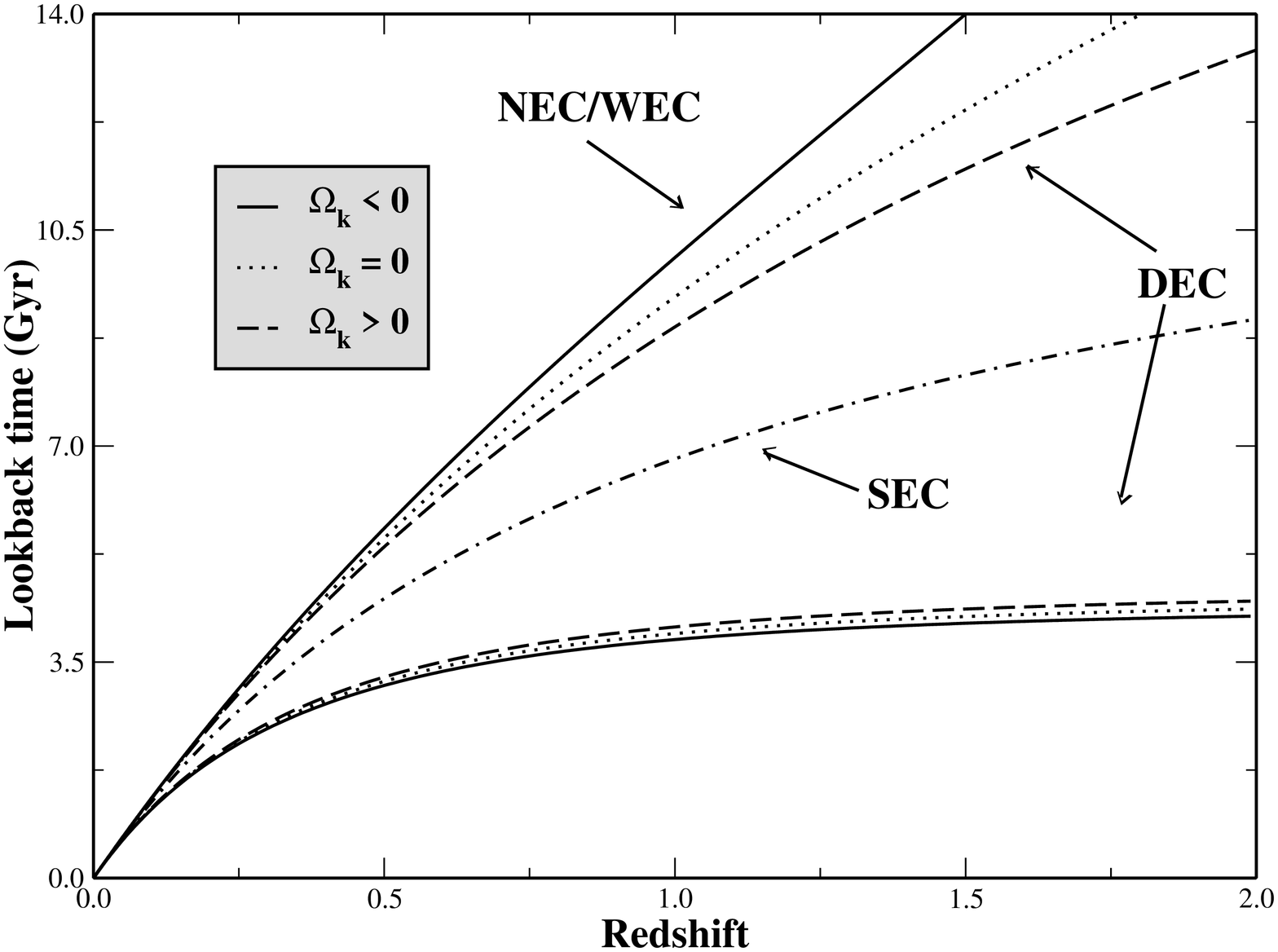}
\includegraphics[width=6.0cm,height=8.0cm,angle=270]{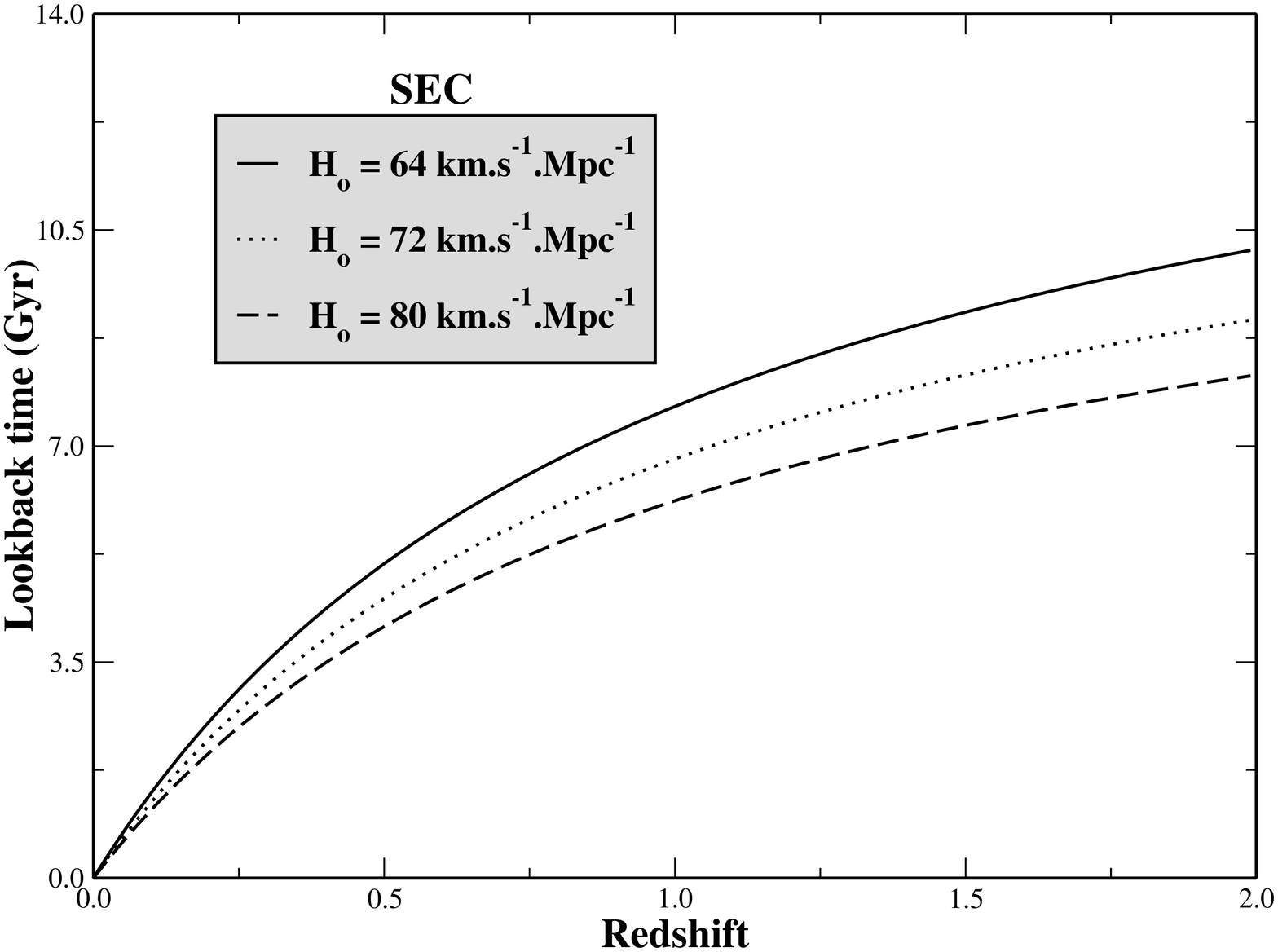}
\caption{Model-independent bounds on the lookback time-redshift relation $t_L(z)$ for different signs of the curvature parameter $\Omega_k$.\ 
{\bf Left}: Upper bounds from the NEC/WEC are shown in the top set of curves, 
while the upper and lower bounds from the DEC correspond, respectively, 
to the indicated top and bottom sets of curves. The SEC upper bound prediction, which 
is curvature-independent [Eq. (\ref{SEC-bound})], is represented by the 
curve in the middle. 
{\bf Right}: The dependence of the SEC predicted upper bound on $t_L(z)$ with the 
Hubble parameter. From top  to botton, the curves correspond to values of $H_0$ 
lying in the $1\sigma$ interval $0.64 \leq H_0 \leq 80$ ($\rm{km \ s^{-1} Mpc^{-1}}$), 
as given by the \emph{HST} key project~\cite{hst}.} 
\end{figure*}

\section{Lookback Time Bounds from Energy Conditions}

The lookback time $t_L$ is defined as the elapsed time between the present age of the universe ($t_0$) and the time $t(z)$, when the light from a cosmic source at a particular redshift $z$ was emitted. In the context of the FLRW models it is given by
\begin{equation}  \label{lookbacktime}
t_L(a)= t_0 - t(z) = \int_a^{a_0} \frac{da}{\dot{a}}\;, 
\end{equation}
where $a_0/a = (1 + z)$ and the subscript 0 denotes present-day 
quantities.

Before deriving the bounds from energy conditions on the lookback 
time-redshift relation, we shall discuss how the lookback time $t_L(z)$ 
can be recast in a suitable form to confront with observational data.  
To this end, following Refs.~\cite{Capozziello-et-al,Pires-et-al,Dantas-et-al}, 
we first note that the age $t_i(z)$ of a cosmic source (a galaxy or a quasar, 
for example) at redshift $z$ is the difference between the age of the Universe 
at $z$ and its age when the galaxy was born (at $z_F$), which in turn can 
be written in terms of the lookback time as 
\begin{equation}
t_i(z)= t_L(z_F) - t_L(z)\;.
\end{equation}
{}From this expression it is straightforward to show that the observed 
lookback time of a cosmic source at $z_i$ can be written as
\begin{eqnarray}  
\label{tauobserved}
t_L^{obs}(z_i) & = & t_L(z_F) - t(z_i) \\  \nonumber 
               & = & t_0^{obs} - t(z_i) - d_f\;,
\end{eqnarray}
where $t_0^{obs}$ is the present estimated age of the Universe and $d_f$ stands for the \emph{incubation time} or \emph{delay factor}, which accounts for our ignorance about the  formation redshift $z_F$ of the source.

\vspace{0.3cm}
\centerline{\bf{A. Bound from the NEC/WEC}} 
\vspace{0.3cm}

In order to obtain the bounds from the NEC/WEC on the lookback time $t_L(z)$, 
we note that for all $a < a_0$ the first integral of Eq.~(\ref{nec-eq}) 
gives
\begin{equation} \label{nec-ineq}
\dot{a} \geq a_0\,H_0 \,\sqrt{\Omega_k+(1-\Omega_k)(a/a_0)^2}\;,
\end{equation}
where $\Omega_k = -k/(a_0 H_0)^2$ and $H_0=\dot{a}(t_0)/a(t_0)$ are, 
respectively, the current values of the curvature and Hubble parameters. 
Now, making use of this inequality, we integrate~(\ref{lookbacktime})  
to obtain the following upper bound for the lookback time:
\begin{widetext}
\begin{eqnarray}  \label{NEC-WEC-bound}
t_L(z) \leq \frac{ H_0^{-1} }{ \sqrt{1-\Omega_k} }
 \left\{ \ln\left[
 \frac{(1+z)}{1 + [1 + ( {\Omega_k \over 1-\Omega_k} )\,(1+z)^2 ]^{1/2}} \right]
   - \ln \left( \frac{ \sqrt{1-\Omega_k} }{1+ \sqrt{1-\Omega_k}} \right) \right\} \;,
\end{eqnarray}
\end{widetext}
which, for $\Omega_k = 0$, reduces to
\begin{equation} \label{necflat}
t_L(z) \leq H_0^{-1} \ln (1+z)\;.
\end{equation}
Concerning the derivation of Eq. (\ref{NEC-WEC-bound}), some words of 
clarification are in order here: 
first, that we have incorporated  the constraint $\Omega_k<1$ that arises 
from the WEC, as given by Eq.~(\ref{wec-eq}); second, by requiring the 
argument of the square root in the first logarithmic term to be positive 
we restrict our analysis of a spatially closed geometry  ($\Omega_k<0$) to 
redshifts lying in the interval $z < \sqrt{|(1- \Omega_k )/\,\Omega_k|}\, -1$. 
Note, however, that given the current estimates of the curvature parameter from 
WMAP and other experiments, i.e., $\Omega_k = -0.014 \pm {0.017}$~\cite{WMAP}, 
the above interval leads to $z \lesssim 9$, which covers  the entire range 
of galaxy observations ($z \lesssim 1.8$) we shall be  concerned with in 
our analysis. Clearly, if the NEC/WEC are obeyed, then $t_L(z)$ must take 
values such that Eq.~(\ref{NEC-WEC-bound}) holds. 

The three curves at the top of Fig.~1(a) illustrate the NEC/WEC bounds on 
the lookback time $t_L$ as a function of the redshift $z$ for different 
signs of the curvature parameter $\Omega_k$. To plot this curves we have 
used the central value of the \emph{Hubble Space Telescope} (\emph{HST}) 
key project estimate, i.e., $H_0 = 72$ $\rm{km \ s^{-1} Mpc^{-1}}$~\cite{hst}. 

\begin{figure*}[t]
\includegraphics[width=6.5cm,height=7.5cm,angle=270]{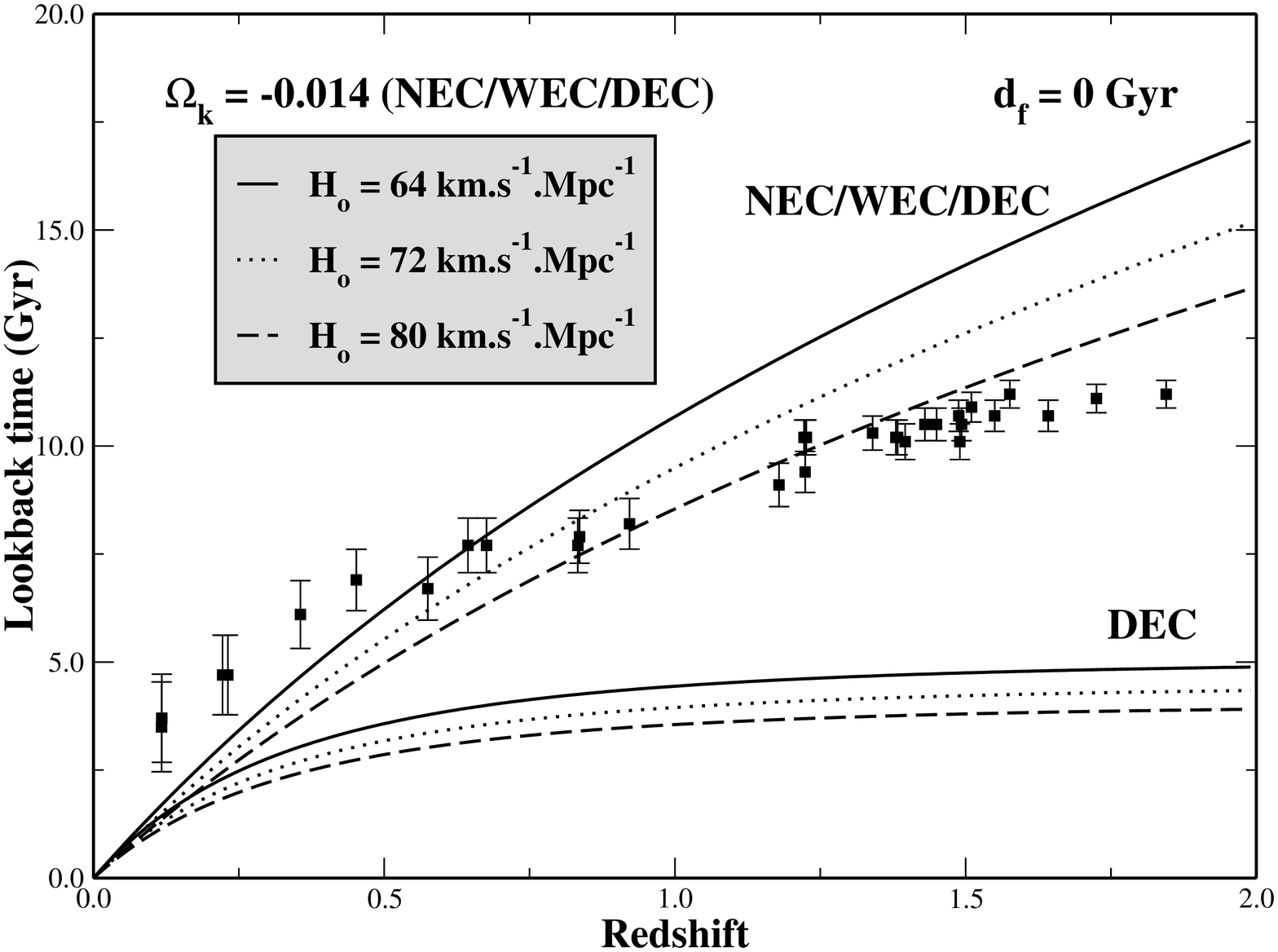}
\includegraphics[width=6.5cm,height=7.5cm,angle=270]{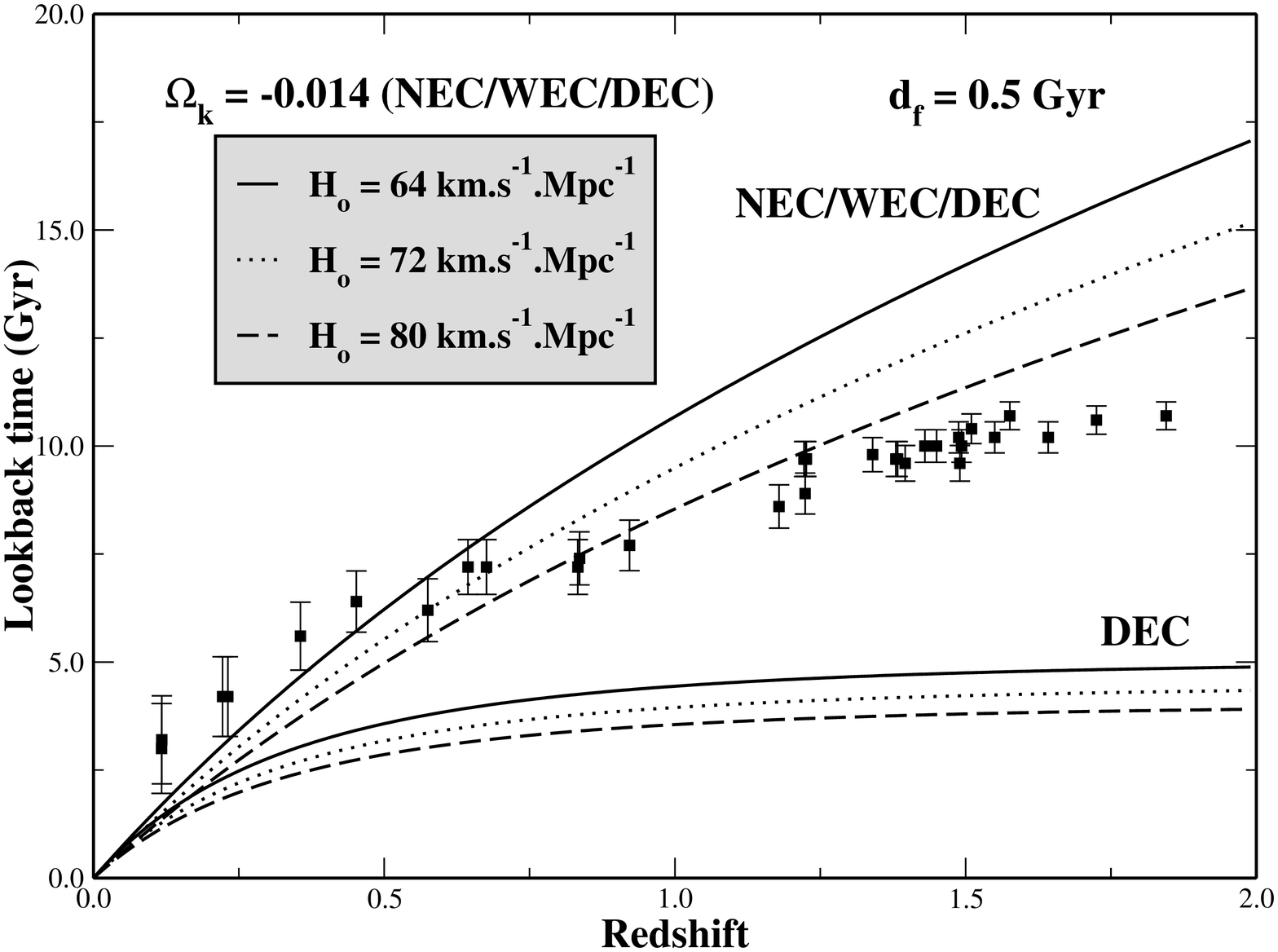}
\caption{Model-independent bounds on $t_L(z)$ as a function of the redshift for different values of the Hubble parameter within the $1\sigma$ interval $H_0 = 72 \pm 8$ $\rm{km \ s^{-1} Mpc^{-1}}$. {\bf Left}: The upper bounds on $t_L(z) $ from the NEC/WEC are shown in the top set of curves, while the upper and lower bounds from the DEC correspond, respectively, to the top and bottom sets of curves as indicated. In this analysis we have fixed $d_f=0$.
{\bf Right}: The same as in the previous panel for $d_f=0.5$ Gyr.} 
\end{figure*}

\vspace{0.4cm}
\centerline{\bf{B. Bound from the SEC}}
\vspace{0.4cm}

Similarly to the NEC/WEC case, the first integral of Eq.~(\ref{sec-eq}) 
clearly gives 
\begin{equation} \label{sec-ineq}
\dot{a} \geq a_0 \, H_0 \;,
\end{equation}
for all $a < a_0$. The above inequality along with equation~(\ref{lookbacktime}) 
furnishes
\begin{equation}  \label{SEC-bound}
t_L(z) \leq H_0^{-1}\,\frac{z}{1+z}\;,
\end{equation}
which, differently from the NEC/WEC case, holds regardless of the value of 
the curvature parameters $\Omega_k$. The dashed-dotted line of Fig.~1(a) shows the upper bound for the SEC-fulfillment prediction [Eq. (\ref{SEC-bound})] 
as a function of the redshift for different values of the Hubble parameters. 

\vspace{0.4cm}
\centerline{\bf{C. Bounds from the DEC}}
\vspace{0.4cm}

From inequation~(\ref{dec-eq}) the DEC provides upper and lower bounds 
on the rate of expansion $\dot{a}$, and therefore gives rise to two 
associated bounds on the lookback time $t_L(z)$. Inequations~(\ref{nec-eq}) 
and~(\ref{wec-eq}) along with~(\ref{dec-eq}) make apparent that the DEC upper 
bound coincides with the NEC/WEC bound given by Eqs.~(\ref{NEC-WEC-bound}) 
and (\ref{necflat}).

Now, in order to set the lower bound from the DEC, we integrate both sides 
of the first inequality~(\ref{dec-eq}) to obtain
\begin{equation}  \label{a-DEC-bound}
\dot a \leq a_0 H_0 \sqrt{ \Omega_k+(1-\Omega_k)(a/a_0)^{-4}} \;.
\end{equation}
Inserting this inequality into the expression~(\ref{lookbacktime}) we obtain 
the following lower bound for the non-flat ($\Omega_k \neq 0$ ) 
FLRW models:
\begin{equation} \label{int18}
t_L(z) \geq {H_0^{-1} \over \sqrt{|\Omega_k| }} \, 
\left(\frac{1-\Omega_k}{|\Omega_k|}\right)^{1/4}
 \int \limits_{x}^{x_0} {{x'}^{2} \; dx' \over
\sqrt{1 \pm {x'}^{4}}} \;,
\end{equation}
where the $\pm$ sign corresponds, respectively, to values of 
$\Omega_k \gtrless 0$, $x =(a/a_0)[|\Omega_k| /(1-\Omega_k)\,]^{1/4}$ 
is a new variable, and the upper limit of integration is 
$x_0=[|\Omega_k| /(1-\Omega_k)]^{1/4}$. Note that the integral~(\ref{int18}) 
can also be expressed in terms of an elliptic integral plus elementary 
functions~\cite{M_Visser1997}. Note also that, for the flat case 
($\Omega_k =0$), Eq.~(\ref{a-DEC-bound}) reduces to 
$\dot a \leq H_0 \;( a_0 ^3 / a^2)$ which, along with~(\ref{lookbacktime}), 
gives the following lower bound on the lookback time
\begin{equation} \label{decflat}
t_L(z) \geq \frac{H_0^{-1}}{3}\,\,[\,1- (1+z)^{-3}] \;. 
\end{equation}
The two sets of curves at the top and bottom of Fig.~1(a) illustrate 
the DEC bounds on the lookback time $t_L$ as a function of the redshift $z$ 
for different signs of the curvature parameter $\Omega_k$. 
Similarly to the previous cases, to plot these curves we have set 
$H_0 = 72$ $\rm{km \ s^{-1} Mpc^{-1}}$.

Before proceeding to our comparison with the observational data, 
it is worth emphasizing that, differently from the case of the 
distance modulus $\mu(z)$ (see Ref.~\cite{SAR2006,SAPR2007}), 
the energy conditions predicted bounds on the lookback time $t_L(z)$ 
depends strongly on the value adopted for the Hubble parameter%
\footnote{As is well known, the uncertainties on the value of the 
Hubble parameter play a very important role on any cosmological 
test involving age estimates. As an example, for the current 
accepted standard scenario, i.e., a flat $\Lambda$CDM model, 
the age of the Universe can be approximated by 
$t_0 \simeq \frac{2}{3}H_0^{-1}\Omega_m^{-0.3}$ Gyr, where $\Omega_m$ 
is the matter density parameter. The error propagation in the determination 
of $t_0$, 
$(\frac{\Delta t_0}{t_0})^2 \simeq (\frac{\Delta H_0}{H_0})^2
+ (0.3\frac{\Delta \Omega_m}{\Omega_m})^2$, 
clearly shows that the fractional error in $H_0$ is three times more important 
than the fractional error in $\Omega_m$~\cite{rees}.}. 
To illustrate this point, we show in Fig.~(1b) the SEC predictions for values of 
$H_0$ lying in the $1\sigma$ interval $H_0 = 72 \pm 8$ $\rm{km \ s^{-1} Mpc^{-1}}$, 
as given by the \emph{HST} key project~\cite{hst}. At $z \simeq 1$, for instance, 
the difference between the $t_L(z)$ SEC prediction for the lower 
($t_L(z) \simeq 7.64$ Gyr) and upper [$t_L(z) \simeq 6.11$ Gyr] limits on $H_0$ 
is $\sim 20\%$. Similar conclusions also apply to the NEC/WEC and DEC lookback 
time predictions.

\begin{figure*}[t]
\includegraphics[width=6.5cm,height=7.5cm,angle=270]{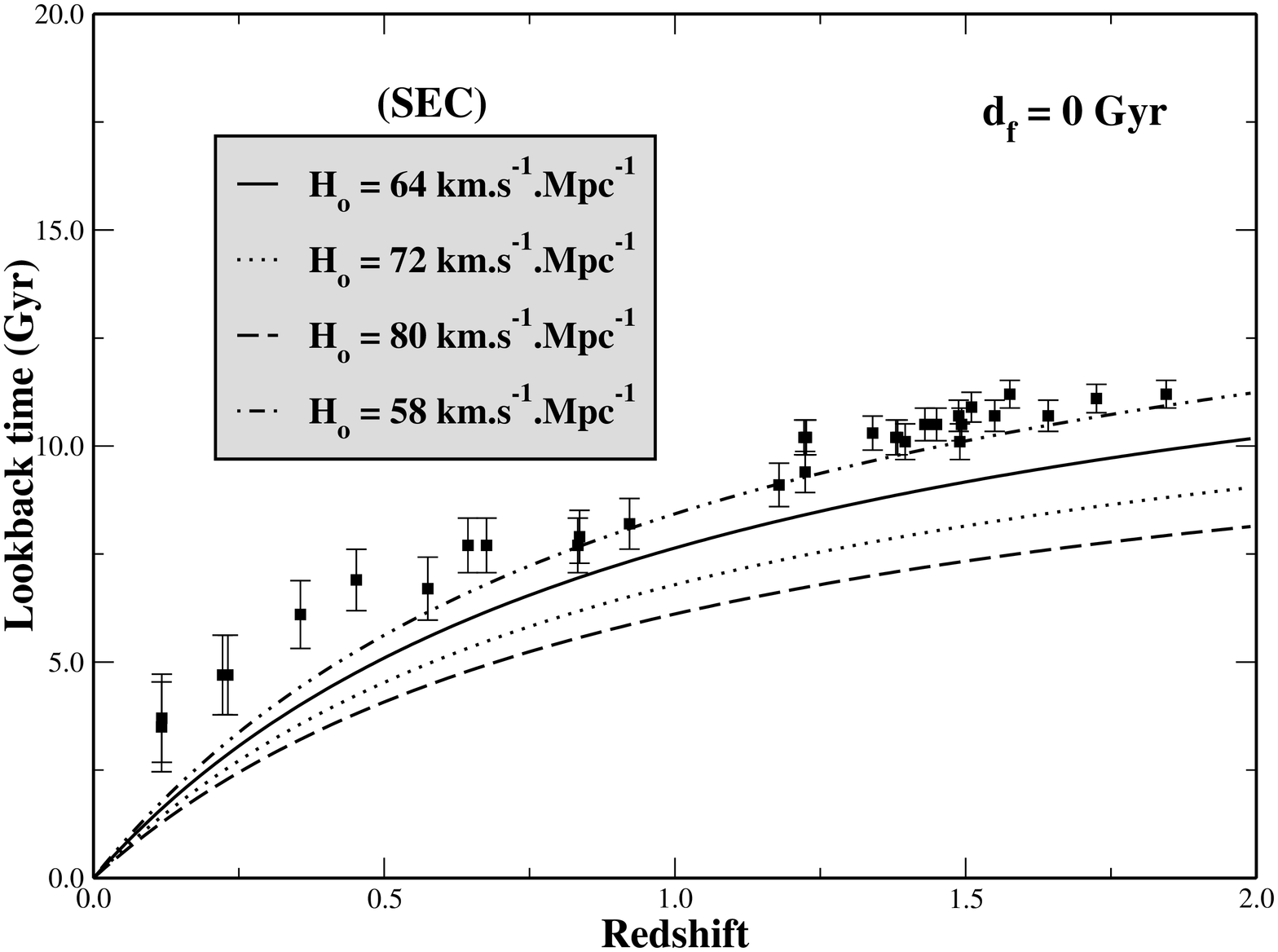}
\includegraphics[width=6.5cm,height=7.5cm,angle=270]{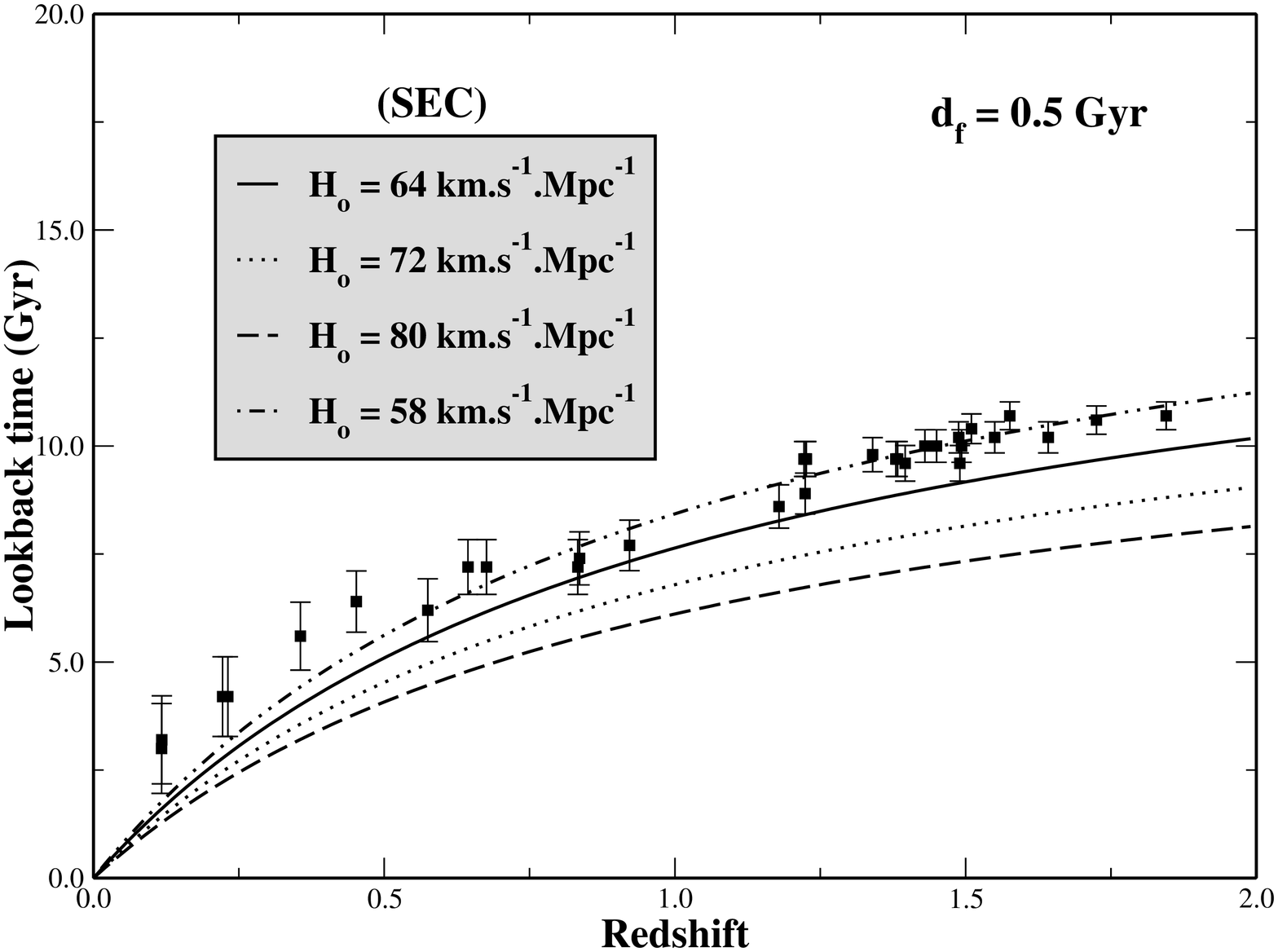}
\caption{
Model-independent bounds from the SEC on $t_L(z)$  as a function of the 
redshift for different values of the Hubble parameter within the $1\sigma$ 
interval $H_0 = 72 \pm 8$ $\rm{km \ s^{-1} Mpc^{-1}}$. 
{\bf Left}: SEC upper bounds on $t_L(z)$ by assuming $d_f = 0$ Gyr. 
{\bf Right}: SEC Upper bounds on $t_L(z)$ by assuming the delay factor to be $d_f = 0.5$ 
Gyr. In both Panels, the SEC-bound curve for the value of the Hubble parameter estimated 
by Sandage and collaborators~\cite{sandage}, i.e., $H_0 = 58$ $\rm{km \ s^{-1} Mpc^{-1}}$ 
is also shown.} 
\end{figure*}

\section{Analysis and Discussion}

\subsection{Data}

We use age estimates of 32 galaxies ranging from $0.11 \lesssim z \lesssim 1.84$, 
as recently analized in Ref.~\cite{Simon_Verde_Jimenez2005}. This sample includes 
objects from the recently released Gemini Deep Deep Survey (GDDS)~\cite{gdds}, 
archival data~\cite{arcdata,dunlop} and the two prototypical evolved red galaxies 
LBDS 59W091 and LBDS 59W069 \cite{dunlop}, whose cosmological impications have 
been largely discussed in the literature~\cite{alcaniz} 
(see~\cite{Simon_Verde_Jimenez2005} for more details). 

In order to build up our lookback time sample,  we combine the ages of the 
above galaxy sample with estimates of the total age of the Universe $t_0^{obs}$, 
according to Eq.~(\ref{tauobserved}). Here, we assume $t_0^{obs} = 13.7 \pm 0.2$, 
as provided by a joint analysis involving current CMB and large-scale 
structure experiments (WMAP, CBI, ACBAR and 2dFGRS)~\cite{WMAP} (see 
also~\cite{tegmark}). The last step toward our lookback time sample concerns 
the delay factor $d_f$ [the third term in the r.h.s. of  Eq.~(\ref{tauobserved})]. 
As discussed in Ref.~\cite{Simon_Verde_Jimenez2005} (see also~\cite{gdds}), the 
most likely star formation history for this galaxy sample is that of a single 
burst of duration less than $0.1$ Gyr, although in some cases the duration of 
the burst is consistent with $0$ Gyr, which means that the galaxies 
have been evolving passively since their initial burst of star formation. 
In the subsequent analyses, to check the influence of $d_f$ on our results, 
we assume following two values for this quantity: $d_f = 0$ Gyr and 
$d_f = 0.5$ Gyr.

\subsection{Results}

In Figs.~(2) and~(3) we confront the energy conditions predictions for 
$t_L(z)$  with current 
lookback time observations for values of the Hubble parameter lying 
in the interval $H_0 = 72 \pm 8$ $\rm{km \ s^{-1} Mpc^{-1}}$ and 
$\Omega_k = -0.014$, which corresponds to the central value of the 
estimates provided by current CMB experiments~\cite{WMAP}.

\subsubsection{NEC/WEC/DEC}

Figure~2(a) shows the upper and lower-bound $t_L(z)$ curves for the 
NEC/WEC and DEC-upper bound fulfillment with $d_f = 0$ Gyr. Similarly to 
the results involving current SNe Ia observations~\cite{SAR2006,SAPR2007}, 
an interesting  aspect of these two panels is that they indicate that 
these energy conditions may have been violated at $z \lesssim 0.5$ 
irrespective of the value of $H_0$ in the above 1$\sigma$ interval. 
A clear example of such violation is given by the galaxies at $z = 0.452$ 
and $z = 0.355$. Indeed, while their observed lookback time are 
$t_L^{obs} = 6.9 \pm 0.71$ Gyr and $t_L^{obs} = 6.1 \pm 0.78$ Gyr, 
the upper-bound NEC/WEC--fulfillment prediction for the corresponding 
redshifts are given, respectively, by $t_L(z = 0.452) = 4.95$ Gyr and 
$t_L(z = 0.355) = 4.02$ Gyr for $H_0 = 72$ $\rm{km \ s^{-1} Mpc^{-1}}$;
and $t_L(z = 0.452) = 5.72$ Gyr and $t_L(z = 0.355) = 4.65$ for  
$H_0 = 64$ $\rm{km \ s^{-1} Mpc^{-1}}$. 
By considering the central value of {\emph{HST}} key project, the discrepancy 
between the observed value and the NEC/WEC--fulfillment 
prediction at $z = 0.452$ is of 1.95 Gyr or, equivalently, $\simeq 2.7 \sigma$, 
which clearly indicates a violation of NEC/WEC at this redshift.%
\footnote{We note that, although the lookback time estimates for the evolved red 
galaxies LBDS 59W091 (at $z = 1.55$) and LBDS 59W069 (at $z = 1.43$)~\cite{dunlop} do not show a clear evidence for violation of the 
NEC/WEC/DEC predictions,  they do for the SEC bounds [Figs. (3a) and (3b)] in the entire interval of $H_0$ considered in this paper.} 
Similar conclusions for the NEC/WEC and DEC (upper bound) are also obtained 
when a delay factor of 0.5 Gyr is considered in our analysis [Fig.~2(b)]. 
In this case, however, a clear evidence for violation of the  
NEC/WEC and DEC (upper-bound) conditions is possible only for values of 
$H_0 > 64$ $\rm{km \ s^{-1} Mpc^{-1}}$. 
For $H_0 = 72$ $\rm{km \ s^{-1} Mpc^{-1}}$, e.g., the difference between 
the  observed values of the lookback time and the NEC/WEC/DEC  
prediction at $z = 0.452$, for instance, is of 1.32 Gyr or, equivalently, 
$\simeq 1.86\sigma$.

Concerning the above results, some interesting aspects are worth 
mentioning at this point.  First, similarly to the results involving 
current SNe Ia observations~\cite{SAR2006,SAPR2007}, all the above 
results holds only for the upper-bound of the DEC predictions, and 
the lower-bound of DEC is not violated by these lookback time data.  
Second, again similarly to the results of Refs.~\cite{SAR2006,SAPR2007}, 
the above analysis is very insensitive to the values of the curvature 
parameter so that all the above conclusions remain unchanged for values 
of $\Omega_k$ within the interval provided by the current CMB experiments, 
i.e., $\Omega_k = -0.014 \pm {0.017}$~\cite{WMAP}.%
\footnote{As an example, by taking the upper and lower 1$\sigma$ limit 
given by WMAP, i.e., $-0.031 \leq \Omega_k \leq 0.003$, the NEC/WEC predicted 
lookback time at $z = 1$ ranges (for $H_0 = 72$ $\rm{km \ s^{-1} Mpc^{-1}}$) 
between $t_L(z = 1) = 9.59$ Gyr and $9.39$ Gyr, respectively, which corresponds 
to a difference of $\simeq 2\%$}
Third, although our analyses and results are model-independent, 
in the context of a FLRW model with a dark energy component parameterized 
by an equation of  state $w \equiv p/\rho$, violation of NEC/WEC and DEC 
is associated with the existence of the so-called \emph{phantom} fields 
($w < -1$), an idea that has been largely explored in the current 
literature~\cite{caldwell}. Therefore, by assuming this standard framework, 
the above results seem to indicate a possible dominion of 
these fields over the conventional matter fields very recently, 
for $z \lesssim 0.5$. 

\subsubsection{SEC}

The upper-bound $t_L(z)$ curves for the SEC-fulfillment are shown in 
Figs.~(3a) and (3b) for different values of the Hubble parameter. 
For $d_f = 0$ Gyr [Fig.~(3a)] and $H_0$ in the 1$\sigma$ interval 
given by the \emph{HST} key project~\cite{hst}, SEC seems to be 
violated in the entire redshift range, without a single galaxy 
in agreement with the theoretical upper-bound SEC prediction. 
Such an outcome is in agreement with both the results of
Ref.~\cite{Stakman-et-al_1999} and the SNe Ia analysis of 
Ref.~\cite{SAPR2007}, although in this latter analysis the first 
clear evidence for SEC violation happens only at $z \simeq 1.2$. 
Interestingly, a better concordance with these SNe Ia results is 
possible when the Hubble parameter is fixed at 
$H_0 = 58$ $\rm{km \ s^{-1} Mpc^{-1}}$, as recently advocated by 
Sandage and collaborators~\cite{sandage} [the upper curve in 
Figs. (3a) and (3b)].
For the central value of {\emph{HST}} key project, the discrepancy 
between the observed value and the SEC-fulfillment prediction value, 
e.g. at $z = 1.64$, is as large as $\simeq 2.26$ Gyr. This means 
that, for $d_f = 0$ Gyr and  by considering the Hubble parameter 
within the current accepted interval, even at very high redshifts, 
i.e., $z \gtrsim 1$, when the Universe is expected to be dominated 
by normal matter, all the lookback time estimates discussed here 
are at least $1\sigma$ higher than the theoretical value derived 
from Eq.~(\ref{SEC-bound}). 
A slightly different conclusion is obtained by considering a delay 
factor of $d_f= 0.5$ Gyr [Fig.~3(b)]. In this case, an interval 
of concordance between the SEC prediction and the observational 
data is possible for $H_0 \lesssim 64$ $\rm{km \ s^{-1} Mpc^{-1}}$, 
which corresponds to the 1$\sigma$ lower bound of the  \emph{HST} 
key project measurements. For its central value, however, although 
reduced relative to the previous case ($d_f = 0$ Gyr), the 
discrepancy between the prediction and observed values for the 
entire interval of redshift is still considerable.

\section{Concluding remarks}

In this work, by using the fact that the classical energy conditions can be recast as a set of differential constraints involving the scale factor $a$ and its derivatives, we have extended and complemented our previous results~\cite{SAR2006,SAPR2007} by deriving model-independent bounds on the lookback time-redshift relation and confronted them with the sample of lookback time measurements discussed in Section III. 
Although the predicted bounds on $t_L(z)$ depend upon the adopted values for the Hubble parameter, we have shown that, similarly to the results involving SNe Ia observations~\cite{SAR2006,SAPR2007}, all the energy conditions seems to have been violated in recent past of cosmic evolution for the current estimated values of $H_0$~\cite{hst} and the value of the curvature parameter in the interval $\Omega_k = -0.014 \pm {0.017}$~\cite{WMAP}. An important outcome of our analyses is that, for the above $H_0$ and $\Omega_k$ intervals, the SEC, whose violation in a FLRW expanding model is closely related to the accelerating expansion of the Universe, seems to have been violated in the entire redshift range of the galaxy sample, i.e., $0.11\lesssim z \lesssim 1.84$. Another interesting aspect related to this SEC violation is that there seem to be a better concordance of the $t_L(z)$ and the SNe Ia results~\cite{SAR2006,SAPR2007} for lower values of the Hubble parameter as, e.g, the one recently advocated by Sandage \emph{et al.}~\cite{sandage}. 

Finally, we emphasize that in agreement with our previous analysis~\cite{SAR2006,SAPR2007} and other recent recent studies~\cite{M_Visser1997}, the results reported in this work reinforce the idea that, in the context of the standard cosmology, no possible combination of  \emph{normal} matter is capable of fitting the current observational data.

\begin{acknowledgments}
J.S. and N.P. are supported by PRONEX (CNPq/FAPERN). J.S. also thanks PCI-CBPF/MCT for the financial support. J.S.A. and M.J.R. thank CNPq for the grants under which this work was carried out. JSA is also supported by FAPERJ No. E-26/171.251/2004.
\end{acknowledgments}

\end{document}